\documentclass[twocolumn,notitlepage,floatfix]{revtex4-2}
\usepackage[english]{babel}
\usepackage[utf8x]{inputenc}
\usepackage{subcaption}

\usepackage{amsmath}
\usepackage{amssymb}
\usepackage{graphicx}

\begin{document}

\title{Bandgaps of bent and buckled carbon nanotubes}

\author{Alex Kleiner \\orcid.org/0000-0002-4694-2218}
\date{January 12, 2023}



\begin{abstract}
Carbon nanotube's large electro-mechanical coupling and robustness makes them attractive for applications where bending and buckling is present. But the nature of this coupling is not well understood. Existing theory treats only  weak and homogeneous deformations. 
 We generalize it and derive close-form expressions for bandgaps under non-homogeneous deformation. The theory is first compared with a number of published DFT simulations, and  then  applied  to the specific case of bending and buckling -- where a kink is present. In the pre-buckling regime, bandgaps change  $\propto\pm \kappa^4$ where $\kappa$ is the bending curvature; inside the kink at post-buckling, while near criticality, the kink is shallow and the gap is $\propto \pm\kappa^{1/2}$, where the sign depends on the chiral integers. For a deeper kink but still with an open  cross-section, both the bandgap  and local Fermi energy strongly downshift.
\end{abstract}
\maketitle

\tableofcontents

\begin{table*}
\begin{center}
\begin{tabular}{l  l} 
 \hline
 \,\,\,Symbol & \,\,\,Definition   \\ [0.5ex] 
 \hline
 \,\,\,$\epsilon$&    axial strain \\ [0.5ex] 
 
  \,\,\,$\zeta$&    axial torsion \\ [0.5ex] 
 
 \,\,\,$\kappa_\theta$ & circumferential curvature   \\[0.5ex] 
 
 \,\,\,$\kappa$ & bending curvature    \\[0.5ex] 
 
 \,\,\,$\kappa_z$ & local axial curvature    \\[0.5ex] 
 
 \,\,\,$\kappa^\textrm{cr}$ & bending curvature at buckling point    \\[0.5ex] 
 
 \,\,\,$b=3.5$ eV/\AA & $d\gamma/dl$, where $l$ is bond length\\
 
 \,\,\,$\gamma=2.7$eV & $\pi$-orbitals overlap energy\\
 
 \,\,\,$n,m$ & nanotube's chiral integers; $m\leq n$\\
 
 \,\,\,$p,q$ & integers; $n-m=3q+p$ (\ref{eq:n-m})\\
 
\,\,\, $v,w$ & in-plain and out-of-plain deformation vectors, respectively (appendix \ref{appendix:brazier})\\
 
\,\,\, $c_h$ & $ac_h$ is the circumference; $c_h=\sqrt{n^2+nm+m^2}$\\
 
 \,\,\,$R$ & nanotube's radius, $2\pi R= ac_h$\\
 
\,\,\, $\nu=0.19$ & Poisson ratio  \cite{yakobson1}; $\nu=0.34$ according to  \cite{poisson2002}; see also \cite{cao}, \cite{like_yakobson} \\
 
 \,\,\,$t=0.66$\AA & effective thickness \cite{yakobson1};  $0.75$\AA \, according to \cite{poisson2002}; see also \cite{cao}, \cite{like_yakobson} \\
 
\,\,\,$\alpha$ & chiral angle; $\cos 3\alpha=(2n^3-2m^3+3n^2m-3nm^2)/2c_h^3$\\
 
 \,\,\,$\xi$ & ovalization parameter (eq. \ref{eq:w})\\
 
 \,\,\,$\xi^\textrm{cr}$ & critical ovalization --  $2/9$ \\[0.5ex] 
 
  \,\,\,$E_g^\pi$ & bandgap due to the $\pi-$band alone \\[0.5ex] 
  
   \,\,\,$E_g^\textrm{pre}$ & bandgap at the pre-buckling regime \\[0.5ex] 
   
    \,\,\,$E_g^\textrm{post}$ & bandgap at the post-buckling regime \\[0.5ex] 
 
 \,\,\,$E_s$ & energy of the singlet-band above graphene's Fermi level \\[0.5ex] 
 
 \,\,\,$C_s$ & coefficient for $E_s$ (appendix \ref{appendix:fermi_and_C_s}); $C_s=8\sim12$eV \AA$^2$\\[0.5ex] 
 
\,\,\,$R_{10,0}$ & the radius of the (10,0) tube; $R_{10,0}=4$\AA\\[0.5ex] 

\,\,\,$R_c$ & upper critical radius; above it, unhybridized $\pi$-band structure is valid\\[0.5ex] 

\,\,\,$R_v$ & lower critical radius; below it, bandgaps are zero\\[0.5ex] 

 \,\,\,$\vec{K}_F$ & Fermi points \\[0.5ex] 
 
  \,\,\,$\delta k_y$ & lateral shift of the Fermi points \\[0.5ex] 
  
   \,\,\,$\vec{l}, l_y, l_z$ & nearest neighbor bond vectors and their lateral and axial projections \\[0.5ex] 
   
    \,\,\,$U$ & elastic energy per unit length \\[0.5ex]

  \hline
\end{tabular}
\end{center}
\caption{List of symbols and values. }
\label{table1}
\end{table*}

\section{Introduction}

Three decades after its discovery, carbon nanotubes has yet to realize its  potential. 
 A slow but continuous progress in fabrication may, however, make feasible a growing number of  application ideas  suggested early-on but never pursued due to economical and fabrication considerations.
 One such application class would be electro-mechanics -- as it relies on the coupling between two of the tube's unique features: one dimensional conductivity and elasticity. 
 
This coupling is most extreme in the kink of a buckled tube. The kink's  size is of the order of the tube diameter, but what is its electronic structure and how will it affect transport as a function of further bending?  Currently, the answer  is far from being clear. The kink is a point of large curvature and non-homogeneous deformation, and, as we show below, non of these is theoretically well understood. 

The elementary theory predicting bandgaps in carbon nanotubes is based on zone-folding of graphene \cite{Jishi1994}. It classifies tubes as metallic or semiconducting   -- with bandgaps $\propto 1/R$ according to their chiral vectors (appendix \ref{appendix:ZF}). 
Corrections $\propto 1/R^2$  due to circumferential curvature \cite{mintmire}\cite{kane}, predict that metallicity is lifted in all but one type: the (achiral) armchair tubes.  
An obvious prediction  is the bandgaps would trend higher at ever smaller radii. But DFT computations (table \ref{table:Eg}) clearly show that the opposite is true: the narrowest tubes are in-fact gapless; it is qualitatively attributed to a curvature-induced hybridization of the $\sigma-\pi$ bands, but it has not been quantitatively incorporated into existing theory. 

This should also play a critical role in the kink, where very large curvature is present.  Thus, our first aim is to understand the effects of large curvature quantitatively. But the other characteristic of the kink -- its large and non-homogeneous deformation, needs also to be accounted for.

The modulation of bandgaps due to \emph{homogeneous} deformation is well-known \cite{yang_deformation}\cite{mypaper1}. But that is not sufficient, even for weak bending without kinks; 
 the strain profile in bending is asymmetric about the neutral line, i.e: pure bending has no \emph{net} strain; this would imply, according to existing theory, no change in the bandgap -- a clear contradiction with simulations \cite{Rochefort}\cite{Rochefort1999}.

Thus, in order to reconcile theory and simulations and  derive closed-form solutions for the bandgap in the kink,  we need, in addition to incorporating effects of large curvature, extend the theory of homogeneous deformation into the non-homogeneous domain. 
 For completeness, however, we start by recapping the existing, homogeneous, theory of bandgaps in carbon nanotubes.

\section{Homogeneous deformation}\label{sec:homogeneous_deformation}
In this section we start by recasting the tight-binding $\pi$-band theory of homogeneous deformation \cite{kane}\cite{yang_deformation}\cite{mypaper1}  in a formalism that will be used here for non-homogeneous fields; we then consider deviations from the pure $\pi$-band, which become dominant at very small radii, or large circumferential curvatures. 

\subsection{General formulation}\label{subsec:homogeneous_general_formulation}
The fundamental theory of bandgaps in carbon nanotubes, reproduced here in appendix \ref{appendix:ZF}, gives  increasingly erroneous gaps as the nanotubes' diameter decreases. The reason lies in a symmetry of graphene that is broken in nanotubes: isotropy of bonds. The assumption that the three nearest overlap integrals are equal (\ref{eq:equal_gamma}), does not fully hold in nanotubes, since neighboring $\pi$-orbitals are not  parallel, nor are neighboring atoms equally separated. 
 Discarding thus assumption (\ref{eq:equal_gamma}), we find the new Fermi points by solving eq. (\ref{eq:wallace_hab_fundamental})
while allowing  small changes to the overlap integrals $\delta\gamma_{j}$. The solution yields  \cite{mypaper1}, 

\begin{eqnarray}\label{eq:delta_kc_app}
    \delta k_y&=&\frac{\textrm{sgn}[1-2p]}{2\pi R\sqrt{3} \gamma}\times\\
&\Big(&(m-n)\delta\gamma_{1}+
    (2n+m)\delta \gamma_{2}-(n+2m)\delta \gamma_{3}\Big).\nonumber
\end{eqnarray}
where $\delta k_y$ is the shift of the Fermi points  along the circumferential coordinate $y$, relative to their zone-folding position (\ref{kfermi}). 
Now,  knowing the $\delta\gamma_j$ in (\ref{eq:delta_kc_app}), translates, through the linear spectrum (\ref{eg1}) to the following gap equation,  
\begin{equation}\label{eq:generic_pi_gap}
    E_g^\pi=\frac{a\gamma}{R}\left|\frac{p}{\sqrt{3}}+\sqrt{3} R\delta k_y\,\right|.
\end{equation}
 The superscript $\pi$ serves to label it as a pure $\pi$-band, in contrast to the hybridized band to be considered later.

\begin{table*}
\begin{center}
\begin{tabular}{ c |   c  c c c c |c } 
 \hline
 \,\,$n,m$\,\, &\,\, ref. \cite{Barone2006}&\,\,\,\, ref. \cite{Shan_prl}\,\,\,& \,\,\,\,\,\,ref. \cite{kurti2003}\,\,\,\,\,\,&\,\,\,\,\,\,ref. \cite{susumu2011}\,\,\,\,\,\,&\,\,\,\,\,\,ref. \cite{kurti2004}\,\,\,\,\,\,&\,\,\,\,\,eqs. (\ref{eq:finall_gap})\,\,\,\,\,\,   \\ [0.5ex] 
 \hline
 
 $4,0$ &   $0$  &   $0$ & $0$&0&0&0\\ [0.5ex] 
 

 $5,0$ &   $0$  &   $0$ & $0$ &0&0&0\\ [0.5ex] 
 
 $5,1$ & - & - &  - &0.05&0.13&0.06\\ [0.5ex] 
 
 $5,3$ &  - & - &  - &1.2&1.18&1.12\\ [0.5ex] 
 
 $5,4$ &  - & - &  - &1.12&-&1.14\\ [0.5ex] 
 
 $6,1$ &  - & - &  - &0.43&0.41&0.71\\ [0.5ex] 
 
 $6,2$ &   - & - &  - &0.7&0.67&0.74\\ [0.5ex] 
 
 $6,4$ &   - & - &   - &1.1&1.09&1.07\\ [0.5ex] 
 
 $7,0$ &  $0.66$  &   $0.34$ & $0.19$ &0.2&0.21&0.474\\ [0.5ex] 
 
 $8,0$ &   $1.05$  &   $0.487$ & $0.73$&0.6&0.59&0.853\\ [0.5ex] 
 
 $10,0$ &   $0.96$  &  $0.87$ & $0.88$&0.8&0.77&0.913\\ [0.5ex] 
 
 $11,0$ &   $1.1$   &  $0.8$ & $1.13$&-&0.93&0.945\\ [0.5ex] 
 
 $13,0$ &   0.79  &  - &0.73 &-&0.64&0.714\\ [0.5ex] 
 
 $14,0$ &  0.85  &  -  & 0.9&-&0.72&0.733\\ [0.5ex] 
 
$16,0$ &   - & 0.57 & 0.61&-&0.54&0.586\\ [0.5ex] 

$17,0$ &   - &  - & -&-&0.58&0.599\\ [0.5ex] 

$19,0$ &    - &  - & -&-&0.46&0.497\\ [0.5ex] 

$20,0$ &   - &  - & -&-&0.5&0.506\\ [0.5ex] 
 \hline
\end{tabular}
\end{center}
\caption{Bandgaps (in eV) of  semiconducting tubes.} 
\label{table:Eg}
\end{table*}

\subsection{Small circumferential  curvature}\label{section:curvature_kane}

One of the symmetries present in graphene but not in nanotubes is the parallel nature of the
$\pi$-orbitals.  In nanotubes, neighboring $\pi$-orbitals are misaligned by a small angle: if $\vec{l}_j$ is the vector connecting them (eqs. \ref{bonds}), and $l_{jy}$ is its $y$-component, the angle is $\beta_j=l_{jy}/R$. Then $\gamma_j\rightarrow\gamma\cos^2\beta_j$ -- which gives
\begin{equation}\label{eq:delta_gamma_curv}
    \delta\gamma_j=-\gamma \frac{l_{jy}^2}{2R^2}.
\end{equation}
Now plugging this in eqs. (\ref{eq:delta_kc_app}) and (\ref{eq:generic_pi_gap}) yields
\begin{equation}\label{eq:curvature_pi_gap}
    \frac{E_g^\pi}{\gamma}=|p|\frac{a}{R\sqrt{3}}+\textrm{sgn}[1-2p]\,\frac{a^2\cos{3\alpha}}{16R^2},
\end{equation}
where $\cos 3\alpha$ is given in table \ref{table1}. 
$E_g^\pi$ includes two effects over the $\pi$-band: the $1/R$-rule of the fundamental gap, and the $1/R^2$ correction due to curvature. Here, in contrast with the fundamental prediction,  ¨metallic¨ tubes ($p=0$)  do open small gaps, and their $1/R^2$ trend, as given by eq. (\ref{eq:curvature_pi_gap}), was experimentally verified on a few large zigzag tubes \cite{science_1_over_Rsquare}.

\begin{figure*}[htbp]  
\begin{center}
\includegraphics[height=0.45\textheight]{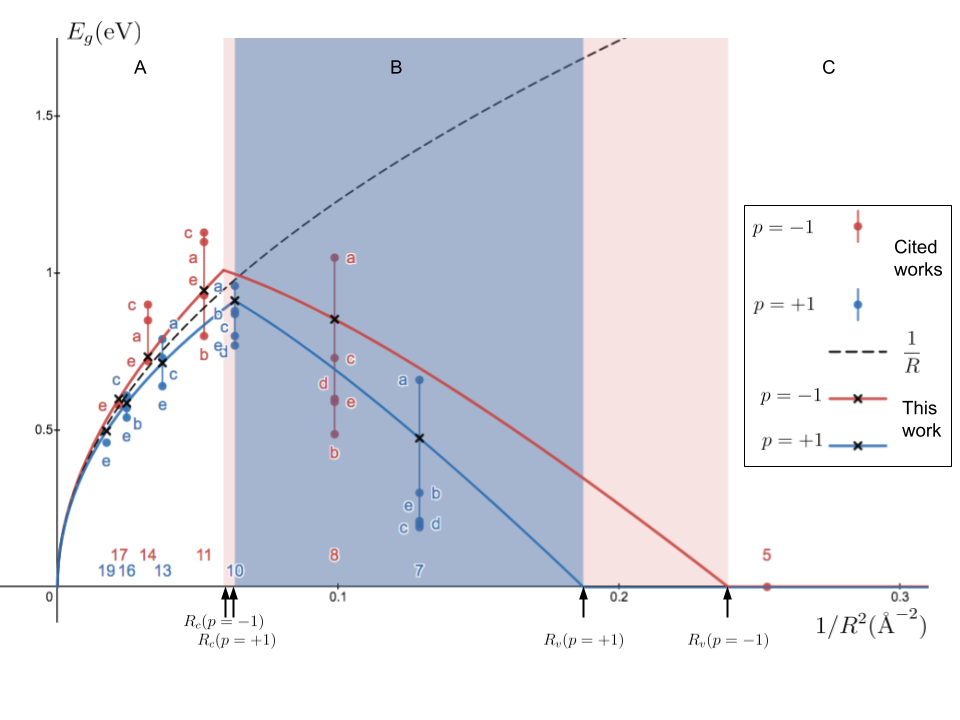}
\caption{Bandgaps of semiconducting zigzag tubes vs. $1/R^2$. The two type of semiconductors, $p=\mp 1$ (for definition see eq. \ref{eq:n-m}), are marked with red and blue colors, respectively.  The  integers above/below the $x$-axis are aliases for the zigzags $(n,0)$; they are positioned under their DFT bandgaps (dots connected by vertical lines); the dots, labels $a,b,c,d$ and $e$, refer to the  correspond publications (see table \ref{table:Eg}). The full curves are plots of eq. (\ref{eq:finall_gap}). The broken curve is the zone-folding "$1/R$-rule" (eq. \ref{eq:Eg_ZF}). It is evident that, for large tubes, this rule holds with slight modification, but for smaller tubes the gap reverses trend and finally vanishes. The critical radius of trend reversal is $\sim 4.2$\AA, which lies between the tubes $(11,0)$ and $(10,0)$. The regions $A$, $B$ and $C$ correspond to the regimes $R\geq R_c$,   $R_v<R<R_c$ and $R\leq R_v$, respectively.
} \label{fig:gap_vs_R_square} 
\end{center}
\end{figure*}

\subsection{Large circumferential  curvature}\label{section:curvature_large}
Predictions from the gap equation (\ref{eq:curvature_pi_gap})  can be compared with MD-simulations: table (\ref{table:Eg}) lists results from simulations run by different groups, and   fig. (\ref{fig:gap_vs_R_square}) compares them with eq. \ref{eq:curvature_pi_gap} for zigzag tubes. 
 Two opposite trends in fig. \ref{fig:gap_vs_R_square} are evident: in tubes larger then (10,0), the gaps seen to  agree with eq. (\ref{eq:curvature_pi_gap}), but smaller tubes reverse this trend and quickly reach metallicity. 

The reason for this  was pointed out early-on  by Blase et al. \cite{blase1994}; it concerns a \emph{singly}-degenerate band (termed below: $S$-band) which is a $\pi^*-\sigma^*$ hybridization.  In graphene, the $S$-band is positioned very high above Fermi level; but  since  hybridization is $\propto 1/R^2$, the $S-$band downshifts with decreasing radius, and   at a certain radius it crosses the conduction band. 
This radius was found \cite{Shan_prl} by DFT to be the radius of the  $(10,0)$ tube.  i.e: in this tube, the $S$ and conduction bands overlap. Its radius is, 
\begin{equation}\label{eq:R_10}
    R_{10,0}\approx4\textrm{\AA}.
\end{equation}

At  smaller radii, the S-band  cuts through the gap (see for example: (8,0) and (7,0) in fig. \ref{fig:gap_vs_R_square}) until, at the smallest radii, it vanishes entirely (see for example: (6,0), (5,0) and (4,0) in table \ref{table:Eg}).

Let $E_s$ be the energy, relative to Fermi level, of the $S-$band at the K-points. Its
$1/R^2$ dependence was found by DFT computations to scale by a  factor of $C_s\sim8$ (ev$\cdot$\AA$^2$) \cite{susumu2011} (see also appendix \ref{appendix:fermi_and_C_s}). But the hybridization near the K-points scales also with chirality $\propto \cos 3\alpha$  \cite{mypaper2}. 
Then, the functional form of $E_s$  is
\begin{equation}\label{eq:Es_def}
    E_s=\frac{1}{2}E_g^\pi(10,0)-C_s\left(\frac{1}{R^2}-\frac{1}{R_{10,0}^2}\right)\cos{3\alpha},
\end{equation}
where $E_g^\pi(10,0)\approx0.93$eV -- is the un-hybridized bandgap of the (10,0) tube.
The radii, $R_c$ and $R_v$, at which the $S-$band cross the conduction and valence bands, respectively, can be found by setting
\begin{eqnarray}
   E_s-\frac{1}{2}E_g^\pi&=0,&\,\,\,\,\,\,\,\,R=R_c,\\
E_g^\pi-C_s\left(\frac{1}{R^2}-\frac{1}{R_{10,0}^2}\right)\cos{3\alpha}&=&0,\,\,\,\,R=R_v.
\end{eqnarray}

This yields
\begin{eqnarray}\label{eq:Rc}
    R_c&=&\frac{1}{2a_c}\left(b_c+\sqrt{b_c^2+4a_cc_c}\right),\,\,\,\textrm{where}\\
    a_c&=&\frac{C_s\cos{3\alpha}}{R_{10,0}^2}+\frac{E_g^\pi(10,0)}{2},\nonumber\\
    b_c&=&|p|\frac{\gamma a}{2\sqrt{3}},\nonumber\\
    c_c&=&\left(C_s+\textrm{sgn}[1-2p]\frac{\gamma a^2}{32}\right)\cos{3\alpha},\nonumber
\end{eqnarray}
and,
\begin{eqnarray}\label{eq:Rv}
    R_v&=&\frac{1}{2a_v}\left(-b_v+\sqrt{b_v^2+4a_vc_v}\right),\,\,\,\textrm{where}\\
    a_v&=&\frac{C_s\cos{3\alpha}}{R_{10,0}^2},\nonumber\\
    b_v&=&|p|\frac{\gamma a}{\sqrt{3}},\nonumber\\
    c_v&=&\left(C_s-\textrm{sgn}[1-2p]\frac{\gamma a^2}{16}\right)\cos{3\alpha},\nonumber
\end{eqnarray}
Eqs. \ref{eq:Rc}, \ref{eq:Rv} yield three pairs of $R_c, R_v$ -- a pair for each $p$. These radii thus delimit the bandgap regimes: large tubes ($R>R_c$) have their bandgap expressed by the usual $\pi$-band, $E_g^\pi$ (eq. \ref{eq:curvature_pi_gap}); smaller tubes ($R_v<R<R_c$) have the $S$-band between the valence and the conduction bands, hence the bandgaps in this regime become
\begin{equation}\label{eq:Ed_pi_menos_C}
    E_g^\pi-C_s\left(\frac{1}{R^2}-\frac{1}{R_c^2}\right)\cos{3\alpha}
\end{equation}
where $E_g^\pi$ is the non-hybridized ($\pi$-band) gap (\ref{eq:curvature_pi_gap})  and $C_S\sim 8$ eV$\cdot$\AA$^2$ is a semi-empirical scaling factor (see appendix \ref{appendix:fermi_and_C_s}).
 The smallest tubes ($R\leq R_v$), on the other hand, are metallic since the $S$-band crossed the valence band. The bandgaps are then given by,
\begin{equation}\label{eq:finall_gap}
E_g=
\begin{cases}
\frac{|p|\gamma a}{R\sqrt{3}}+\textrm{sgn}[1-2p]\frac{\gamma a^2}{16R^2}\cos{3\alpha},\,\,\,\,\,\,\,\,R\geq R_c,&\\
\frac{|p|\gamma a}{R\sqrt{3}}+\left(\textrm{sgn}[1-2p]\frac{\gamma a^2}{16}-C_s\right)\frac{\cos{3\alpha}}{R^2}+&\\ +\frac{C_s\cos{3\alpha}}{R_c^2},  \,\,\,\,\,\,\,\,\,\,\,\,\,\,\,\,\,\,\,\,\,\,\,\,\,\,\,\,\,\,\,\,\,\,\,\,\,\,\,\,\,\,\,\,\,\,\,\,\,\,\,R_v<R<R_c,& \\
0,\,\,\,\,\,\,\,\,\,\,\,\,\,\,\,\,\,\,\,\,\,\,\,\,\,\,\,\,\,\,\,\,\,\,\,\,\,\,\,\,\,\,\,\,\,\,\,\,\,\, \,\,\,\,\,\,\,\,\,\,\,\,\,\,\,\,\,\,\,\,\,\,\,\,\,\,\,\,\,\,\,\,\,  R\leq R_v.&
\end{cases}
\end{equation}
This is the complete solution for bandgaps under no deformation. It gives comparable results with various published simulations (table \ref{table:Eg});  fig. (\ref{fig:gap_vs_R_square}) plots the bandgaps of zigzag tubes according to eqs. (\ref{eq:finall_gap}) and the  simulations. 



\begin{figure*}[htbp]  
\begin{center}
\includegraphics[height=0.45\textheight]{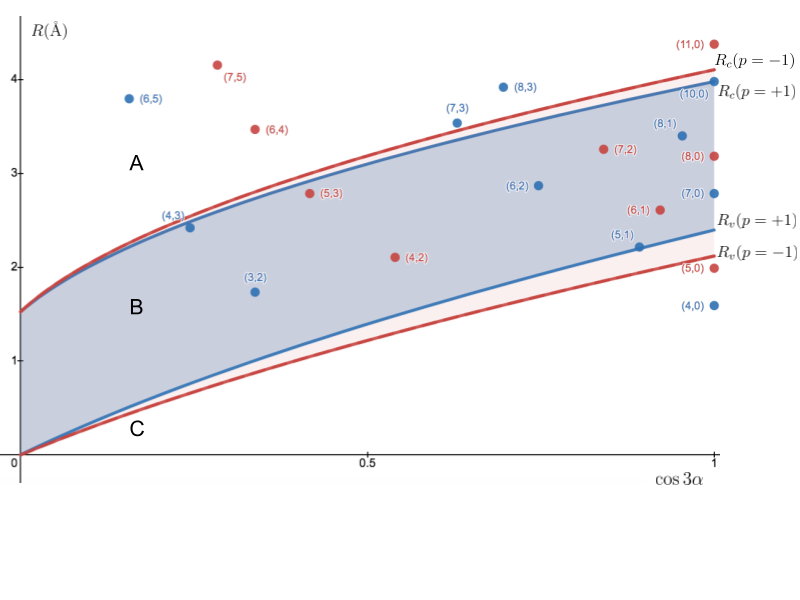}\caption{$R_c$ and $R_v$ (eqs. \ref{eq:Rc}, \ref{eq:Rv}) for $p=+1$ (red) and $p=-1$ (blue) tubes. The bandgap of tubes whose radius lies above $R_c$ is not affected by hybridization. While the gap of tubes that lie between $R_c$ and $R_v$ (colored background) is affected -- and reduced. The couple of tubes lying below $R_v$ have zero gap. The regions $A$, $B$ and $C$ , as in fig. (\ref{fig:map_alpha_r}), correspond to the regimes $R\geq R_c$,   $R_v<R<R_c$ and $R\leq R_v$, respectively.
} \label{fig:map_alpha_r} 
\end{center}
\end{figure*}

\subsection{Strain}
Having found the general gap equations (\ref{eq:finall_gap}), we wish to find its response to homogeneous deformation, such as axial strain, $\epsilon_z$. Following the procedure in \cite{mypaper1} and \cite{yang_deformation}, the bond vectors (\ref{bonds})  $\vec{l}_j=(l_{jy},l_{jz})$, transform as 
\begin{equation}\label{eq:general_bond_vector_deformation}
    \vec{l}_j\rightarrow {\bf D}\vec{l}_j,
\end{equation} 
where ${\bf D}$ is a deformation matrix, given,  for axial strain, by
\begin{equation}\label{eq:strain_D}
    {\bf D}=
    \begin{pmatrix}
1-\nu\epsilon_z & 0\\
0 & 1+\epsilon_z 
\end{pmatrix}
\end{equation}
where $\nu$ is the tube's Poisson ratio (table \ref{table1}). This deformation alters the bond distances, $|l_j|\rightarrow |l_j|+\delta |l_j|$, which in turn, alters the overlap integrals $\gamma_j\rightarrow \gamma_j+\delta \gamma_j\approx \gamma_j+ b\delta|l_j|$, where $b\approx 3.5$eV/\AA (\cite{mypaper1}). This yields, using (\ref{eq:general_bond_vector_deformation}) and (\ref{eq:strain_D}), 
\begin{equation}\label{eq:delta_gamma_strain}
    \delta\gamma_j=\frac{\epsilon_zb\sqrt{3}}{a}(-\nu l_{jy}^2+l_{jz}^2).
\end{equation}
Now plugging the above three $\delta\gamma_j$s in eq. (\ref{eq:delta_kc_app}), we get
\begin{equation}
    \delta k_y=\frac{1}{2\gamma}\textrm{sgn}[2p-1]
\epsilon_z b(1+\nu)\cos{3\alpha}. 
\end{equation}
This equation is the explicit form of eq. (\ref{eq:delta_kc_app}) under axial strain. On substitution in eq. (\ref{eq:generic_pi_gap}) we get,

 \begin{eqnarray}\label{eq:Eg_strain}
     E_g^\pi=\frac{|p|\gamma a}{R\sqrt{3}}+ &\left(\frac{\gamma a^2}{8R^2}-\epsilon_z (1+\nu)ba\sqrt{3}\right)&\nonumber\\
      &\times \frac{1}{2}\textrm{sgn}[1-2p]\cos{3\alpha}&
\end{eqnarray}
 But as discussed above, at small radii the $\pi$-band is crossed by and hybridized by the singlet band, downshifting the gap (\ref{eq:Ed_pi_menos_C}), until it vanishes. This yields the three regimes, as in eqs. \ref{eq:finall_gap},
\begin{equation}\label{eq:full_strain_gap}
E_g(\epsilon_z)=
\begin{cases}
E_g^\pi,\,\,\,\,\,\,\,\,\,\,\,\,\,\,\,\,\,\,\, \,\,\,\,\,\,\,\,\,\,\,\,\,\, \,\,\,\,\,\,\,\,\,\,\,\,\,\,\,\,\,\,\,\,\,\,\,\,\,\,\,\,\,\,\,\,\,\,\,\,\,\,\,\,\,\,\,\,\,  R\geq R_c,&\\
E_g^\pi-C_s\left(\frac{1}{R^2}-\frac{1}{R_c^2}\right)\cos{3\alpha},\,\,R_v<R<R_c& \\
0,\,\,\,\,\,\,\,\,\,\,\,\,\,\,\,\,\,\,\,\,\,\,\,\,\,\,\,\,\,\,\,\,\,\,\,\,\,\,\,\,\,\,\,\,\,\,\,\,\,\,\,\,\,\,\,\,\,\,\,\,\,\,\,\,\,\, \,\,\,\,\,\,\,\,\,\,\,\,\,\,\,\,\,   R\leq R_v.&
\end{cases}
\end{equation}
An unequivocal experimental verification of eqs. (\ref{eq:full_strain_gap}) is not known to us, but the strain and chiral angle dependence  of $E_g^\pi$  was probed by \cite{dai_prl} and qualitatively confirmed. Simulations of compressive   strain in zigzag tubes \cite{umeno2019} appear to agree with both the linear behaviour and $p$-dependence of $E_g^\pi$ as given by eq. (\ref{eq:Eg_strain}).

\subsection{Torsion}
To find the bandgap under an axial torsion, $\zeta$, we start by setting its deformation matrix,
\begin{equation}\label{eq:torsion_D}
    {\bf D}=
    \begin{pmatrix}
1 & 0\\
-\zeta & 1
\end{pmatrix}.
\end{equation}
following the  steps applied to  strain  (\ref{eq:strain_D}-\ref{eq:full_strain_gap}), we get for the un-hybridized $\pi$-bandgap
\begin{eqnarray}\label{eq:Eg_torsion}
     E_g^\pi=&\frac{|p| \gamma a}{R\sqrt{3}} + \frac{1}{2}\textrm{sgn}[1-2p]\times\nonumber\\
     &\left(\frac{\gamma a^2\cos{3\alpha}}{8R^2}-\zeta b a \sqrt{3}\sin{3\alpha}\right).
\end{eqnarray}
As with strain, this bandgap is valid only in the regime where the $S$-band is above the ($\pi$) conduction band ($R>R_c$, eq. \ref{eq:Rc}). Otherwise, hybridization with the $S$-band, as discussed above, strongly downshifts it. In complete equivalence to strain (\ref{eq:full_strain_gap}), the torsion bandgaps yield the general formulae,

\begin{equation}\label{eq:full_torsion_gap}
E_g(\zeta)=
\begin{cases}
E_g^\pi,\,\,\,\,\,\,\,\,\,\,\,\,\,\,\,\,\,\,\, \,\,\,\,\,\,\,\,\,\,\,\,\,\, \,\,\,\,\,\,\,\,\,\,\,\,\,\,\,\,\,\,\,\,\,\,\,\,\,\,\,\,\,\,\,\,\,\,\,\,\,\,\,\,\,\,\,\,\,\,  R\geq R_c,&\\
E_g^\pi-C_s\left(\frac{1}{R^2}-\frac{1}{R_c^2}\right)\cos{3\alpha},\,\,R_v<R<R_c,& \\
0,\,\,\,\,\,\,\,\,\,\,\,\,\,\,\,\,\,\,\,\,\,\,\,\,\,\,\,\,\,\,\,\,\,\,\,\,\,\,\,\,\,\,\,\,\,\,\,\,\,\,\,\,\,\,\,\,\,\,\,\,\,\,\,\,\,\, \,\,\,\,\,\,\,\,\,\,\,\,\,\,\,\,\,   R\leq R_v,&
\end{cases}
\end{equation}
 where $E_g^\pi$ is the torsion-modified bandgap (\ref{eq:Eg_torsion}) of the  $\pi$-band alone.  

\section{Non-homogeneous deformation}\label{subsec:non_homogeneous_general_formulation}
So far, the bandgaps found here concerned tubes with circular cross-section and axial uniformity. But that is not the case in many situations. The cross-section of a bent tube, for example, is compressed in the inner side of bending-curvature and stretched in the other. Its deformation profile is thus varying throughout the unit cell of the tube. 
To treat this and similar cases, we need to generalize the formalism  of section (\ref{subsec:homogeneous_general_formulation}) to include  these non-homogeneous deformation.

\subsection{General formulation}

In the theoretical treatment of section \ref{subsec:homogeneous_general_formulation}, it was tacitly assumed that 
the deformation  is everywhere identical. This allowed us to deform a single graphene unit cell and extract the shift of the Fermi points from there. But that can not be done when the deformation varies
 throughout the tube's unit cell.
Hence, if $N$ is the number of graphene unit cells in the nanotube unit cell, the position of the Fermi points, normally found by solving eq. (\ref{eq:wallace_hab_fundamental}), is now found by  the sum  over the $N$\, $A$-atoms
\begin{equation}\label{eq:new_wallace}
    \sum_{A=1}^N\sum_{j=1}^3 \gamma_{Aj} e^{i\vec{k}\cdot\vec{l}_{Aj}}=0.
\end{equation}

As before, we seek the lateral shift $\Delta K_y$ which is a simple sum of $\delta k_y (A)$ (\ref{eq:delta_kc_app}) of the constituent $A$-atoms,
\begin{equation}\label{eq:Dky1}
    \Delta K_y=\frac{1}{N}\sum_{A=1}^N \delta k_y(A)
\end{equation}
where $\delta k_y(A)$ is now a function of the local strain and curvature at the position of the corresponding $A$'th atom.

The overlap integrals $\gamma_{Aj}$ in eq. \ref{eq:new_wallace} are functions of the in-plain  and out-of-plain deformations of the bond vectors $\vec{l}_{Aj}$. In-plain deformation, such as strain and torsion, changes the $\gamma$'s by changing the distance between neighboring $\pi$-orbitals,  while  out-of-plain deformation, such as the tube's curvature, lowers the $\gamma$'s by having  
the  orbitals misaligned.
  
  In this section we include both deformations without assuming their homogeneity throughout the tube.
  First  consider curvature.  On a plain cutting through the cross section, let $\pm\phi_y$ be the angles between neighboring $\pi$ orbitals and the normal passing through the middle point between them;    and let $\pm\phi_z$ be the corresponding angles   projected on a plain along the axis. If  $\kappa_\theta,\kappa_z$ are the  coordinate curvatures defined as (radius of curvature)$^{-1}$ within the respective plains, and
  writing bond vectors (\ref{bonds}) in the form $\vec{l}=l_y\hat{y}+l_z\hat{z}$, one gets  $|\phi_y|=|l_y|\kappa_\theta/2$ and $|\phi_z|=|l_z|\kappa_z/2$.
     The overlap integral becomes $\gamma\rightarrow\gamma\cos\phi_y\cos\phi_z$, so that for small change $\delta\gamma$  is given by
\begin{equation}\label{eq:delta_gamma_curvature}
    \delta\gamma^\textrm{curv}=-\frac{\gamma}{8}\left(l_y^2\kappa_\theta^2 + l_z^2\kappa_z^2\right).
\end{equation}

Let us now include local axial strain. Its deformation matrix, given by eq. (\ref{eq:strain_D}), induces a change $\delta\gamma^\textrm{strain}$ according to eq. (\ref{eq:delta_gamma_strain}). 
 The total change in the local overlap integrals is then, 
\begin{equation}\label{eq:delta_gamma_tot}
    \delta\gamma=\delta\gamma^\textrm{curv}+\delta\gamma^\textrm{strain}\equiv l_y^2D_y+l_z^2D_z,
\end{equation}
where
\begin{eqnarray}
    D_y&=&\frac{\nu b \sqrt{3}}{a}\epsilon_z - \frac{\gamma}{8}\kappa_\theta^2,\label{eq:D_y}\\
    D_z&=&-\frac{b \sqrt{3}}{a}\epsilon_z - \frac{\gamma}{8}\kappa_z^2.\label{eq:D_z}
\end{eqnarray}
Substituting $l_y, l_z$ from eqs. \ref{bonds} in eq. (\ref{eq:delta_gamma_tot}) we find,

\begin{eqnarray}\label{eq:delta_gamma_3}
\delta\gamma_{A1}&=&\frac{a^2}{4c_h^2}\left((n+m)^2D_y+\frac{(n-m)^2}{3}D_z\right)\nonumber\\
    \delta\gamma_{A2}&=&\frac{a^2}{4c_h^2}\left(m^2D_y+\frac{(2n+m)^2}{3}D_z\right)\\
    \delta\gamma_{A3}&=&\frac{a^2}{4c_h^2}\left(n^2D_y+\frac{(2m+n)^2}{3}D_z\right)\nonumber
\end{eqnarray}
Inserting now eqs. \ref{eq:delta_gamma_3} in \ref{eq:delta_kc_app} 
\begin{equation}
    \delta k_y=\textrm{sgn}[1-2p]\frac{a (D_z-D_y)}{2\sqrt{3}\gamma}\cos{3\alpha}.
\end{equation}
This, by eq. (\ref{eq:Dky1}), we sum over all $A$-atoms in the tube's unit cell,
\begin{equation}\label{eq:Dky2}
    \Delta K_y=\textrm{sgn}[1-2p]\frac{a\cos{3\alpha}} {2\sqrt{3}\gamma N}\sum_{A=1}^N (D_z-D_y). 
\end{equation}
Converting now the sum to an integral,
\begin{equation}\label{eq:Dky3}
    \Delta K_y=\textrm{sgn}[1-2p]\frac{a\cos{3\alpha}} {2\sqrt{3}\pi\gamma }\int_{0}^{2\pi} (D_z-D_y)d\theta. 
\end{equation}
Once $\Delta k_y$ is known, the energy gap follows immediately from the dispersion relation (\ref{eg1}) and the zone-folding gap (\ref{eq:Eg_ZF}),
\begin{equation}\label{eq:generic_non_homo_pi_gap}
    E_g^\pi=\frac{a\gamma}{R}\left|\frac{p}{\sqrt{3}}+\sqrt{3} R\Delta k_y\,\right|.
\end{equation}

 Equation (\ref{eq:generic_non_homo_pi_gap}) is the non-homogeneous version of eq. (\ref{eq:generic_pi_gap}), and likewise, it includes only the $\pi$-band. But now the non-homogeneity of the deformation must be included through the integration in (\ref{eq:Dky3}),   we thus replace $1/R^2$ in eq. (\ref{eq:Es_def}) accordingly,
 \begin{equation}\label{eq:replace_curv_by_int}
     \frac{1}{R^2}\rightarrow \frac{1}{2\pi}\int_0^{2\pi}\kappa_\theta^2\, d\theta.
 \end{equation}
 
Now the full set of gap equations for non-homogeneous deformation can be obtained by using equations (\ref{eq:finall_gap}) with the substitution of eq. (\ref{eq:generic_non_homo_pi_gap}) for  $E_g^\pi$, and a replacement as in  (\ref{eq:replace_curv_by_int}).

\subsection{Elastic theory of bending and buckling}\label{section:elastic_theory}

The structural properties of bending and buckling of SWCNT's had been
 simulated \cite{yakobson1}\cite{like_yakobson}\cite{ijima}\cite{wang}\cite{postbuckling_salerno} \cite{state_art}  and experimented \cite{ijima}\cite{poisson2002} in the years following nanotubes'  discovery. They established that  SWCNT's abide by continuous elasticity theory given an effective   wall ¨thickness¨ of $0.66$\AA\, (values by other groups in table \ref{table1}). Which is always much smaller then the tube's diameter.
 Hence SWCNTs are also  ¨shells¨; precisely: \emph{slender cylindrical shells}.
 
 The elastic theory of their bending, up to the onset of buckling, was developed a century ago by Brazier \cite{brazier}.  It is summarized in appendix \ref{appendix:brazier} for convenience. 
 
 Brazier's fundamental insight was that, in order to reduce shear  under bending,  the tube's circular cross-section  is ovalized  (fig. \ref{fig:brazier}).  At the critical bending, however,  the elastic cost of increased ovalization exceeds the reduction in shear energy -- and the tube buckles.   
 
 In the pre-buckling regime,  Brazier's theory gives the exact shape of the ovalized cross-section, parametrized by $\xi$ in fig. \ref{fig:brazier}. It also predicts, as reproduced in the appendix, that  at the onset of buckling the ovalization is $\xi=2/9$  (eq. \ref{eq:xi_final_app}), independent of Poisson's ratio or other material properties.

At the post-buckling regime, the elastic energy associated with ovalization (beyond the buckling point)is 
\begin{equation}\label{eq:Delta_U_xi_square}
    \Delta U=G(\xi-\xi^\textrm{cr})^2,\,\,\,\,\,\,\,\,\,\,\,\,\,\,\xi\geq\xi^\textrm{cr}
\end{equation}
where $G$ is a constant. 
But MD simulations \cite{yakobson1}\cite{ijima} showed that in this regime, the post-buckling energy density is linear with bending curvature, that is
\begin{equation}\label{eq:U}
    \Delta U=Q (\kappa-\kappa^\textrm{cr})  ,\,\,\,\,\,\,\,\,\,\,\,\,\kappa\geq\kappa^\textrm{cr},
\end{equation}
where $Q$ is a constant and $\kappa^\textrm{cr}$ is the critical bending curvature (\ref{eq:kappa_buckle}). 
Comparing eqs. (\ref{eq:Delta_U_xi_square})  and (\ref{eq:U}),
\begin{equation}\label{eq:zeta_square_root}
    \xi-\xi^\textrm{cr}=\frac{Q}{G} (\kappa-\kappa^\textrm{cr})^{1/2} ,\,\,\,\,\,\,\,\,\,\,\,\,\kappa\geq\kappa^\textrm{cr}.
\end{equation}
This gives the proportionality relation \emph{in the kink}, between bending curvature, $\kappa$, and ovalization of the cross-section , $\xi$. Outside of the kink, however, 
  This  implies that outside the kink, strain is independent of further bending -- its energy is absorbed in the kink, which  acts as a  hinge between the two fixed sections of the tube 
  \cite{mamalis}\cite{mechanism2004}.
 
 A realistic closed-form model of such kinks is not known to us. But simulations revealed \cite{transient_deformation} that for bending not  too far from the onset of buckling, where the kink is shallow,  the cross-section is shaped as an oval, just as the oval parametrized by Brazier in the pre-buckling regime (appendix \ref{appendix:brazier}).
 Since this regime is an intermidiate stage between  the onset of buckling and a fully developed kink, it is also called the \emph{transient-regime} \cite{transient_deformation}. 
 
 This regime ends when the opposite walls of the kink approach contact; at this point the distance between the walls is comparable with the inter-planar distance in graphite $d_g\sim3.35$\AA.  
 In Brazier's parametrization this distance  is  $2R(1-\xi)$ (fig. \ref{fig:brazier}). We thus  have
\begin{equation}
\label{eq:xi_close}
    \xi^\textrm{close}=1-\frac{d_g}{2R},
\end{equation}
where the superscript \emph{close} signifies the  closing of the cross-section in the kink and bringing the opposite walls to near contact. $\xi^\textrm{close}$ marks the upper limit of flattening within which our analysis is expected to hold. For a tube of 1nm in diameter, $\xi^\textrm{close}\sim2/3$.

 The cross-section of a kink in the transient regime is thus bounded by the onset of buckling, at $\xi=2/9$, and $\xi^\textrm{close}$, having the curvature dependence according to eq. (\ref{eq:zeta_square_root}),  namely, 
\begin{eqnarray}
\label{eq:xi_kink}
    \xi^\textrm{kink}=\frac{2}{9}+\left(\frac{\kappa-\kappa^\textrm{cr}}{\kappa^\textrm{close}-\kappa^\textrm{cr}}\right)^{1/2}\left(\xi^\textrm{close}-\frac{2}{9}\right),\\
    \textrm{where}\,\,\kappa^\textrm{cr}\le\kappa\le\kappa^\textrm{close}.\nonumber
\end{eqnarray}
Equation \ref{eq:xi_kink}  demonstrates the evolution of $\xi^\textrm{kink}$, being continuous at the onset of buckling  $\xi^\textrm{kink}=2/9$, through further bending  where $\xi^\textrm{kink}\propto\kappa^{1/2}$, and finally reaching the closing point at $\xi^\textrm{close}$. 

Having found $\xi^\textrm{kink}$, the circumferential curvature in the center of the kink is given, as in the pre-buckling regime, through the out-of-plane deformation $w$ (eq. \ref{eq:w}) and eq. (\ref{eq:kappa_b}) by,
\begin{equation}\label{eq:kappa_kink}
     \kappa_\theta^\textrm{kink}=\frac{1}{R}(1+3\xi^\textrm{kink}\cos{2\theta}).
\end{equation}

The axial strain under bending, $\epsilon_z$, is anti-symmetric with respect to the neutral plane; it thus has no contribution to the integration in eq. (\ref{eq:Dky3}); also the bending curvature, $\kappa_z\ll 1/R$, can be neglected in the integration (\ref{eq:Dky3}), although, as will be shown next, it  affects $\kappa_\theta$. Hence we set in eqs. (\ref{eq:D_y}-\ref{eq:D_z}) $D_y=-\frac{\gamma}{8}\kappa_\theta^2$ and $D_z=0$.  

According to the elastic theory of bending (appendix \ref{appendix:brazier}), the circular cross-section of a bent tube becomes increasingly oval with bending. Parametrizing the ovalization by  $\xi$ (fig. \ref{fig:brazier}), let the circumferential integral of  the square of the curvature (eq. \ref{eq:kappa_b}) be $I$, then
\begin{equation}\label{eq:curv_square1}
    I\equiv\frac{1}{2\pi}\int_0^{2\pi} \kappa_\theta^2 d\theta =\frac{1+\frac{9}{2}\xi^2}{R^2},
\end{equation}
 where the pre-buckling regime corresponds to $0\leq\xi\leq 2/9$. The point $\xi=2/9$ corresponds to the critical curvature at buckling (eq. \ref{eq:xi_final_app}).

\subsection{Bending -- pre-buckling}

Since the ovalization in this regime is a quadratic function of the bending curvature (eq. \ref{eq:xi_ff}), the replacement rule (\ref{eq:replace_curv_by_int}) becomes,
\begin{equation}\label{eq:replace_R_by_kappa}
   I= \frac{1}{R^2}\rightarrow \frac{1+\frac{9}{2}\xi^2}{R^2} = \frac{1}{R^2}\left(1+(L\kappa)^4\right).
\end{equation}
where 
\begin{equation*}
    L\equiv (1-\nu^2)^{1/2} \left(\frac{9}{2}\right)^{1/4} \frac{R^2}{t}
\end{equation*}
where $\nu$ and $t$ are given in table \ref{table1}.
With this replacement, eqs. (\ref{eq:finall_gap}) yield the explicit gap equations under bending,
\begin{equation}\label{eq:pre_A}
    E_g^\textrm{pre}(A)=\frac{|p|a\gamma}{R\sqrt{3}}+\frac{\textrm{sgn}[1-2p]\,\gamma a^2}{16R^2} \left(1+(L\kappa)^4\right)
  \cos{3\alpha},
\end{equation}
where the superscript "pre" refers to pre-buckling and "$A$" corresponds to $R\geq R_c$. Region $B$ then corresponds to radii in the range $R_v<R<R_c$; the bandgaps there are,
\begin{eqnarray}\label{eq:pre_B}
    E_g^\textrm{pre}(B)=\frac{|p|a\gamma}{R\sqrt{3}}&+&\frac{(\textrm{sgn}[1-2p]\,\gamma a^2-16C_s)}{16R^2}\nonumber\\ 
    &\times& \left(1+(L\kappa)^4\right) \cos{3\alpha},
\end{eqnarray}
and finally , at the smallest radii range, $C$, where $R\leq R_v$,
\begin{equation}\label{eq:pre_C}
    E_g^\textrm{pre}(C)=0.
\end{equation}

Eqs. (\ref{eq:pre_A}--\ref{eq:pre_C}) give the bandgaps for all radii in the pre-buckling regime; the three ranges of radii, $A,B$ and $C$ are also shown in  figs. (\ref{fig:gap_vs_R_square}) and (\ref{fig:map_alpha_r}). These equations reveal that, depending on the sign of $p$, bending  may increase, decrease or even close the gap -- a prediction which also coincides with
simulations of ovalized cross-sections \cite{barboza}\cite{Shan_apl2005}.

\begin{figure}[!]  
\begin{center}
\includegraphics[width=0.45\textwidth]{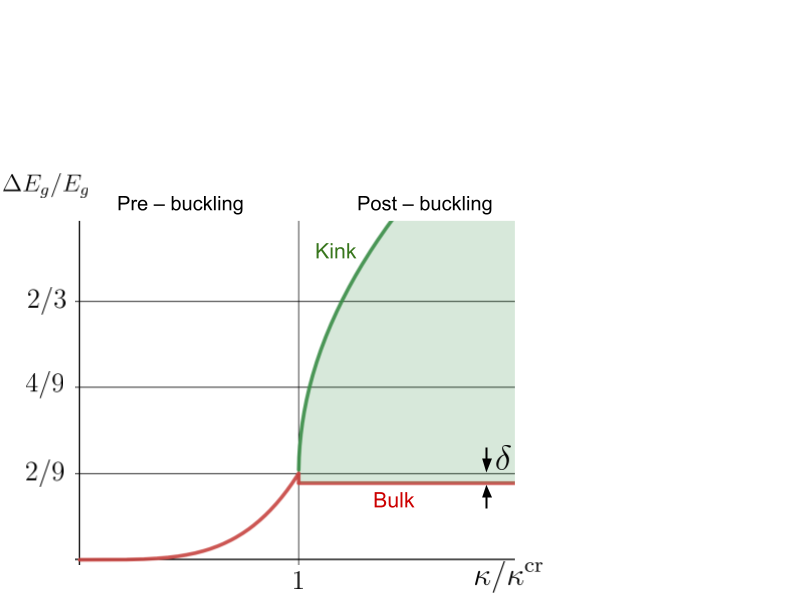}\caption{
Bandgap vs. bending curvature for primary metallic tubes, $p=0$, (armchairs excluded).  $E_g$ is the gap of the straight tube (eq. \ref{eq:curvature_pi_gap}); $\Delta E_g$ is the additional gap due to pure bending (\ref{eq:pre_A}-\ref{eq:pre_C}) and (\ref{eq:post_A}-\ref{eq:post_C}); $\kappa$ is the bending curvature and $\kappa^\textrm{cr}$ is the critical curvature (eq. \ref{eq:kappa_buckle}). In the pre-buckling regime the gap $\propto \kappa^4$  up to buckling point where it is higher by a factor of $2/9$ compared with the straight state (eq. \ref{eq:critical_gap_non_homo_metallic}). At post-buckling, the bulk relaxes  by a small amount $\delta$ (\ref{eq:delta}) and remains there, while in the kink it is  $\propto (\kappa-\kappa^\textrm{cr})^{1/2}$.
} 
\label{fig:metallic_spec} 
\end{center}
\end{figure}


\begin{figure*}[!]
\centering
\begin{subfigure}{.49\textwidth}
  \centering
  \includegraphics[width=0.9\linewidth]{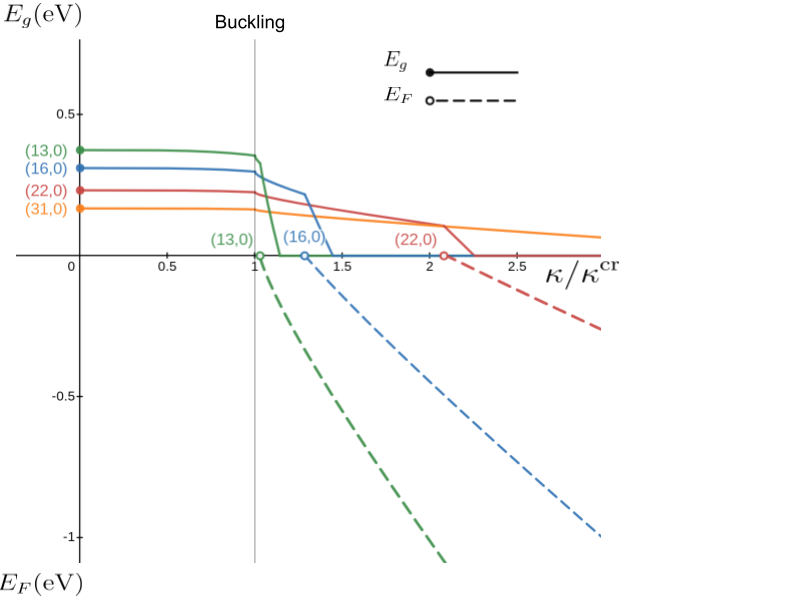}
  \caption{$p=+1$}
\end{subfigure}%
\begin{subfigure}{.49\textwidth}
  \centering
  \includegraphics[width=0.9\linewidth]{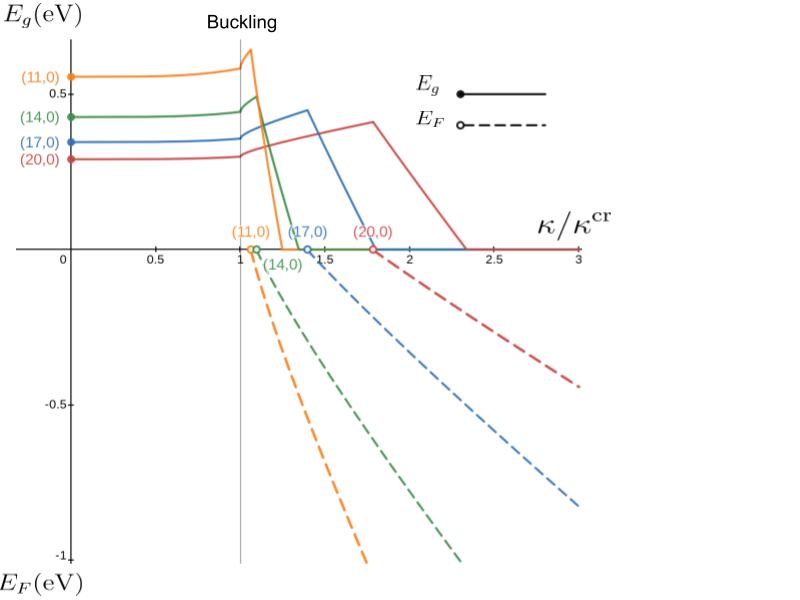}
  \caption{$p=-1$}
\end{subfigure}
\caption{Bandgaps and Fermi energies vs. bending for  semiconducting zigzag tubes of the two types ($p=\pm1$). Following eq. (\ref{eq:pre_A}), the initial gaps  first  decrease (left) / increase (right) $\propto \kappa^4$; at post-buckling it evolves faster: $\propto (\kappa-\kappa^\textrm{cr})^{1/2}$; at increased curvature -- where the  singlet band crosses the conduction band -- the gap strongly downshifts in all cases (\ref{eq:post_B}); at this point, also the Fermi energy (\ref{eq:EF2})  downshifts. The point of $E_F=0$ in the plots corresponds to the value of $E_F$ in graphene.}
\label{fig:gaps_and_Fermi}
\end{figure*}

\subsection{Bending -- critical curvature}
At the critical point of buckling the ovalization parameter $\xi^\textrm{cr}=2/9$ (appendix \ref{appendix:brazier}). This gives for the rhs of eq. (\ref{eq:curv_square1})  $I=\frac{11}{9R^2}$. It is interesting to note that for metallic tubes ($p=0)$, replacing  the rhs of eq. \ref{eq:replace_R_by_kappa} for this value of $I$ and substituting  in eq. \ref{eq:pre_A}, gives,
 \begin{equation}\label{eq:critical_gap_non_homo_metallic}
\frac{\Delta E_g(\kappa=\kappa^\textrm{cr})}{E_g(\kappa=0)}=\frac{2}{9},\,\,\,\,\,\,\,\,\,\,\,\,\,\,\,\,\,\,\,\,p=0,\,\,R\geq R_c.
\end{equation}

This universal ratio for metallic tubes (armchair tubes excluded as their $E_g=0$), relates their bandgaps at the critical point of buckling with their straight value.
Figure (\ref{fig:metallic_spec}) depicts this in terms of the change in the bandgap as a function of the bending curvature.

\subsection{Bending -- post-buckling}
Since the elastic bending moment reaches maximum at buckling point \cite{shell_book}, the buckling transition is accompanied by a small relaxation of the strain energy \cite{yakobson1} \cite{ijima} outside of the kink. This relaxation also lowers Brazier's ovalization and, as will be shown next, also the associated bandgap.

The pre-buckling elastic energy per unit length (eq. \ref{eq:total_energy})  can be approximated near critical point by $U\sim \frac{1}{2}R^3tE\pi\kappa^2$ .
 Denoting the post-buckling elastic energy relaxation per unit length by $\Delta U$, the relaxed state corresponds to a lower curvature given by $\delta\kappa=\Delta U/(\partial U/\partial\kappa)$. The associated shift in the band-gap ratio $\Delta E_g/E_g$ for metallic tubes (eq. \ref{eq:critical_gap_non_homo_metallic}) can be found by expanding this ratio near on the pre-buckling side of criticality, where the function is well behaved $\delta(\Delta E_g/E_g)=\partial(\Delta E_g/E_g)/\partial\kappa)\delta\kappa$ where the derivative is taken at the buckling curvature $\kappa=\kappa^\textrm{cr}$  (eq. \ref{eq:kappa_buckle}); this yields,

 \begin{equation}\label{eq:delta}
     \delta\left(\frac{\Delta E_g}{E_g}\right)=\frac{R \Delta U}{3\pi D},
 \end{equation}
 where $D$ is the elastic rigidity (eq. \ref{eq:D_def}) and $\Delta U$ is the elastic energy  relaxation per unit length at buckling. 
Equation \ref{eq:delta} links, for metallic tubes, the mechanical post-buckling relaxation step with the post-buckling electronic band-gap change in the bulk. 
In contrast with the relaxation in the bulk, the bandgap in the kink is now a stronger function of bending curvature $\kappa$; this can be seen by applying the replacement rule (\ref{eq:replace_curv_by_int}) on the gap eqs.  (\ref{eq:finall_gap})  using $\kappa_\theta^\textrm{kink}$ (\ref{eq:kappa_kink}). 

Staying, as we do throughout this work, within the transient regime where the kink remains open ($\xi^\textrm{kink}<\xi^\textrm{close}$, eqs. \ref{eq:xi_close}, \ref{eq:xi_kink}), bandgaps in the kink fall in three regimes;  these are related to the three regimes in straight tubes (fig. \ref{fig:gap_vs_R_square}), corresponding to whether the gap is determined by the $\pi$-band alone (regime $A$), the modified regime where the singlet ($S$) band hybridizes with the $\pi$ band (regime $B$), or zero -- where the $S$-band downshifted enough to completely close the gap (regime $C$). All tubes fall into one of these regimes depending on their diameter (largest to smallest -- regimes $A$ to $C$, respectively).  Here, however, the regimes are determined by the curvature within the kink which is bounded by $I<1/R_c^2$ (regime $A$), $1/R_v^2>I>1/R_c^2$ (regime $B$), and $I>1/R_v^2$ (regime $C$), where $I$ is given by eq. (\ref{eq:curv_square1}) and $R_c$, $R_v$ are given by eqs. (\ref{eq:Rc} -- \ref{eq:Rv}), respectively. 

Explicitly, regime $B$ is bounded by
\begin{equation}
    \frac{2}{9}\left(\frac{R^2}{R_c^2}-1\right)<(\xi^\textrm{kink})^2< \frac{2}{9}\left(\frac{R^2}{R_v^2}-1\right),
\end{equation}
while regime $A$ applies at the lower bound and regime $C$ in the upper.

The bandgaps of the kink in the various regimes of bending can now be found by  replacing $1/R^2$ in eqs. (\ref{eq:finall_gap}) with $(1+\frac{9}{2}(\xi^\textrm{kink})^2)/R^2$ (eq. \ref{eq:replace_R_by_kappa}), which gives 

\begin{eqnarray}\label{eq:post_A}
     E_g^\textrm{kink}(A)=\frac{|p|a\gamma}{R\sqrt{3}}&+&\frac{\gamma a^2}{16R^2}\textrm{sgn}[1-2p]\,\nonumber\\
     &\times&\left(1+\frac{9}{2}(\xi^\textrm{kink})^2\right)
  \cos{3\alpha},
\end{eqnarray}
where  $\xi^\textrm{kink}$ is given by eq. (\ref{eq:xi_kink}). In region $B$,  eqs. (\ref{eq:finall_gap}) then give,
\begin{eqnarray}\label{eq:post_B}
    E_g^\textrm{kink}(B)=\frac{|p|a\gamma}{R\sqrt{3}}&+&\frac{(\textrm{sgn}[1-2p]\,\gamma a^2-16C_s)}{16R^2}\nonumber\\ 
    &\times&\left(1+\frac{9}{2}(\xi^\textrm{kink})^2\right) \cos{3\alpha},
\end{eqnarray}
and as before , in regime $C$,
\begin{equation}\label{eq:post_C}
    E_g^\textrm{kink}(C)=0.
\end{equation}

Eqs. (\ref{eq:post_A}--\ref{eq:post_C}) give the bandgaps in the center of the kink for the respective radii ranges. Here, as in pre-buckling regime, the bandgap as a function of bending can increase, decrease, or vanish, depending on the sign of $p$. The difference with the pre-buckling regime, however, is that here it is $\propto(\kappa-\kappa^\textrm{cr})^{1/2}$; this behaviour is depicted for zigzag tubes where $p=+1$ (fig. \ref{fig:gaps_and_Fermi}a) and $p=-1$ (fig. \ref{fig:gaps_and_Fermi}b) and for metallic tubes of all chiral angles in fig. (\ref{fig:metallic_spec}).

It may be useful to compare the bandgaps of the kink with the bulk (far from the kink -- where $|z|\gg R$).
Assuming that in the bulk $\xi$ remains near criticality, the bandgap in regime $B$ (eq. \ref{eq:post_B}) compared with the bulk gives,
\begin{equation}\label{eq:Eg_diff_kink_bulk}
    E_g^\textrm{kink}-E_g^\textrm{bulk} = -\frac{9C_s}{2R^2} \left((\xi^\textrm{kink})^2-\left(\frac{2}{9}\right)^2\right)\cos{3\alpha},
\end{equation}
where $\xi^\textrm{kink}$ is given by eq. (\ref{eq:xi_kink}). 

It is worth noting that for armchair tubes ($p=0, \alpha=30^o)$, these equations (as well as the pre-buckling ones) predict no bandgaps. That holds, however, as long as the underlying assumption through this work -- that the kink remains open \cite{transient_deformation} -- holds.

 \section{Summary}
 
 We presented here a comprehensive theory of  bandgaps in carbon nanotubes, including strong curvature and  large non-homogeneous deformation. This theory reconciles the fundamental theory of bandgaps in nanotubes (appendix \ref{appendix:ZF}) and its well-known corrections  (at small deformation and small curvature),   with  contrasting results  from a host of DFT computations of very small or  highly deformed tubes; the present theory shows them to be  special cases of the same general equations (\ref{eq:finall_gap}).
 
 A formalism was derived to calculate the gaps due to  a general non-homogeneous circumferential deformation (by starting with eq. \ref{eq:finall_gap} and making the replacement \ref{eq:replace_curv_by_int}).
 We then  applied this formalism to study bending, including buckling and a kink (with a caveat of staying within the transient  regime, i.e: where the opposite walls of the kink do  not touch).
 
 The results detail the  gap evolution under both weak and strong bending.  In the pre-buckling regime, the bandgap shifts   $\propto \pm \kappa^4$. A notable result is that, by the onset of buckling, the gaps of primary-metallic tubes, independent of chiral vector or radius, increase by a ratio of $2/9$ compared with their un-bent value,  (eq. \ref{eq:critical_gap_non_homo_metallic} and fig. \ref{fig:metallic_spec}).
 
 In the post-buckling regime ($\kappa>\kappa^\textrm{cr}$, eq. \ref{eq:kappa_buckle}), at first, the bandgap in the kink  shifts  $\propto \pm (\kappa-\kappa^\textrm{cr})^{1/2}$, up-to the point where the singlet band crosses the conduction $\pi$-band initiating a  steep downshift to zero (eqs. \ref{eq:post_A}-\ref{eq:post_C}, and fig. \ref{fig:gaps_and_Fermi}). The downshift in the bandgap is accompanied, in this regime, by a substantial downshift of the Fermi energy (appendix \ref{appendix:fermi_and_C_s}). 


\section{Appendices}

\appendix
\section{Brazier's Theory}\label{appendix:brazier}

\begin{figure*}[htbp]
\centering
\begin{subfigure}{.5\textwidth}
  \centering
  \includegraphics[width=0.8\linewidth]{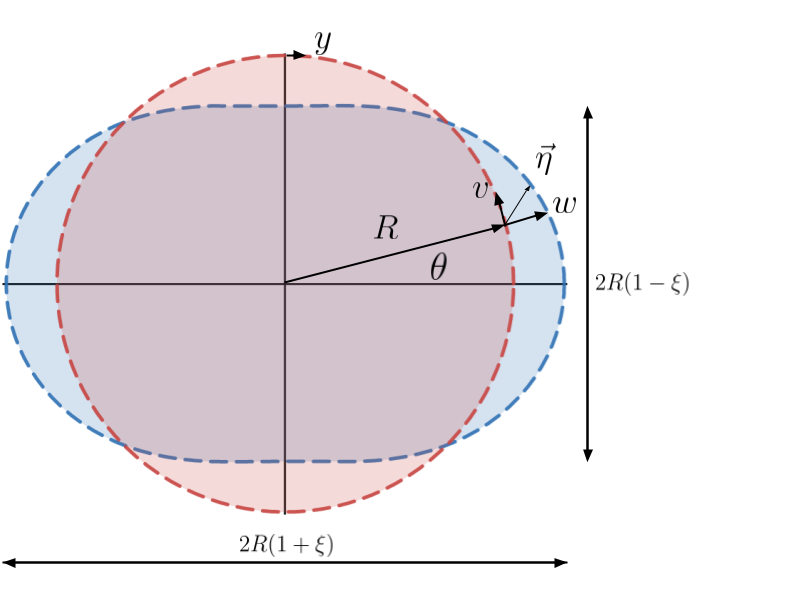}
  \caption{Cross-section}
\end{subfigure}%
\begin{subfigure}{.5\textwidth}
  \centering
  \includegraphics[width=0.8\linewidth]{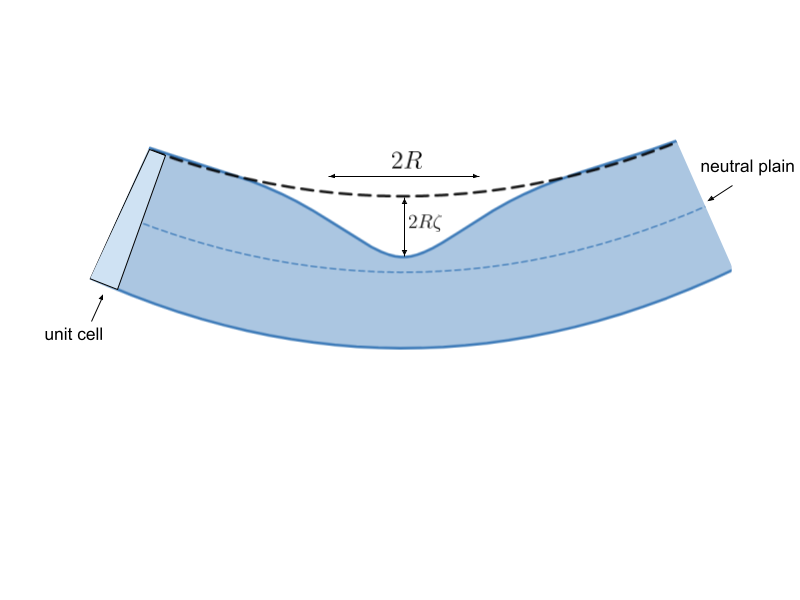}
  \caption{Profile}
\end{subfigure}
\caption{Bending causes the circular cross-section to become oval, turning  $R\rightarrow R(1+\xi\cos{2\theta})$ where $\xi$ is the ovalization parameter; at $\xi=2/9$ (blue shape) the tube reaches criticality and buckles. }
\label{fig:brazier}
\end{figure*}




This appendix has a number of  relevant derivations from the elastic theory of thin cylindrical shells under pure bending. The theory was derived by Brazier \cite{brazier}. For  the exposition we follow reference   \cite{shell_book}.

Consider a slender cylindrical shell, initially straight, under bending.  The  strain profile  is anti-symmetric about the neutral plane, compressive in the inner side and tensile in the outer;  the energy per unit length  is given by
\begin{equation}\label{eq:U_s1}
    U_\textrm{z}=\frac{1}{2}I \kappa^2
\end{equation}
where $\kappa$ is the bending  curvature and $I$ is the second moment of area; for a perfectly circular cross-section 
\begin{equation}\label{eq:I0}
    I\equiv I_0=\pi R^3t.   
\end{equation}
where $t$ is the thickness of the tube.

 It will be shown below that by flattening the cross section the tube reduces its energy.  It is reasonable to assume that small flattening  can be expressed by an  out-of-plain displacement  of the form

\begin{equation}\label{eq:w}
    w=R\xi\cos{2\theta}.
\end{equation}
where $\xi$ (see fig. \ref{fig:brazier}) is a dimensionless measure of the flattening, or ovalization. The curvature due to ovalization is
\begin{equation}\label{eq:kappa_b}
    \delta\kappa_\theta=-\frac{\partial^2w}{\partial y^2}- \frac{w}{R^2}=\frac{3\xi}{R}\cos{2\theta},
\end{equation}
where $y=R\theta$ is the circumferential coordinate of the original circle, the second derivative is the usual curvature due to local changes of displacement while the second component is due to the change of radius. The in-plain deformation $v$, together with $w$ (\ref{eq:w}), determines the circumferential strain $\epsilon_y=\partial v/\partial_y + w/R$. As a first approximation Brazier assumed the surface to be inextentional, ie: $\epsilon_y=0$, yielding 
\begin{equation}\label{eq:v}
    v=-\frac{R}{2}\xi\sin{2\theta}.
\end{equation}
where we assumed no net rotation by taking the constant of integration to be zero. It may be commented that for the electronic structure calculation we take a much looser requirement, instead of assuming the surface to be inextentional it is assumed only that the circumference is unchanged; i.e:  $\oint\epsilon_ydy=0$.

The second moment of area in eq. \ref{eq:U_s1} is actually defined as
\begin{equation}\label{eq:Imoment}
    I=\int_0^{2\pi}R t (R\sin\theta+\eta)^2 d\theta,
\end{equation}
where $t$ is the surface thickness and $\eta$ is given by
\begin{equation}\label{eq:eta}
    \eta=w\sin\theta+v\cos\theta=-R\xi\sin^3\theta
\end{equation}
where the last step used eqs. \ref{eq:w} and \ref{eq:v}. Inserting this in eq. \ref{eq:Imoment} we finally get
\begin{equation}
    I=I_0\left(1-\frac{3}{2}\xi+\frac{5}{8}\xi^2\right)
\end{equation}
where $I_0$ is given by \ref{eq:I0}. 
The strain energy per unit length (eq. \ref{eq:U_s1}) is then given by
\begin{equation}
    U_\textrm{z}=\frac{1}{2}\kappa^2 R^3 t E\pi  \left(1-\frac{3}{2}\xi+\left[\frac{5}{8}\xi^2\right]\right)
\end{equation}
where $E$ is the Young's modulus; the $\xi^2$ component in the square brackets is truncated in Brazier's analysis.

The energy per unit length due to Brazier deformation is 
\begin{equation}\label{eq:U_B1}
    U_B=\frac{1}{2}D\int_0^{2\pi}\delta\kappa_\theta^2 R d\theta,
\end{equation}
where $\delta\kappa_\theta$ is given by eq. \ref{eq:kappa_b} and $D$ is the elastic rigidity given by 
\begin{equation}\label{eq:D_def}
    D=\frac{Et^3}{12(1-\nu^2)}.
\end{equation}
Integration of eq. \ref{eq:U_B1} gives
 \begin{equation}
    U_B=\frac{3\pi E t^3}{8(1-\nu^2)}\frac{\xi^2}{R}.
\end{equation}

The total elastic energy per unit length is
\begin{equation}\label{eq:U_tot1}
    U=U_\textrm{z}+U_\textrm{B}.
\end{equation}
Now $\xi$ can be found by requiring $\partial U/\partial\xi = 0$.   The result is
\begin{equation}\label{eq:xi_ff}
\xi=\frac{(1-\nu^2)R^4}{t^2}\kappa^2.
\end{equation}
The values of $\nu$ and $t$ were found by a number of groups (see table \ref{table1}), so we can write
\begin{equation}\label{eq:xi_B}
    \xi=BR^4\kappa^2,
\end{equation}
where $B=2.2$(\AA$^{-2}$) by the values of \cite{yakobson1} compared with $B=1.57$(\AA$^{-2}$) by \cite{poisson2002}.

The total energy (eq. \ref{eq:U_tot1}) is finally given by
\begin{equation}\label{eq:total_energy}
    U=\frac{1}{2}R^3tE\pi \kappa^2 - \left(\frac{3R^7(1-\nu^2)E\pi}{8t}\right)\kappa^4
\end{equation}

The tube begins to buckle when the bending moment $M=d U/d\kappa$, reaches maximum; hence the curvature at buckling point can be found by putting $d^2U/d\kappa^2=0$, which gives
\begin{equation}\label{eq:kappa_buckle}
    \kappa^\textrm{cr}=\frac{t}{3R^2}\sqrt{\frac{2}{1-\nu^2}}.
\end{equation}
$\kappa^\textrm{cr}$ is the critical curvature at the onset of buckling. 
 Using the same values as for eq. \ref{eq:xi_B}, we find
\begin{equation}\label{eq:cr_A2Rsquare}
    \kappa^\textrm{cr}=\frac{A}{R^2},
\end{equation}
where $A=0.316$\AA\, using the values of \cite{yakobson1}, compared with $A=0.376$\AA\, by \cite{poisson2002}.
  MD simulations confirmed the general shape eq. (\ref{eq:cr_A2Rsquare}) and found $A$ to range between  $0.185$\,\cite{cao} and $0.387$\AA\, \cite{yakobson1}  Regardless of the actual numerical value of $A$, substituting \ref{eq:kappa_buckle} in \ref{eq:xi_ff} gives the exact ovalization parameter at buckling point
\begin{equation}\label{eq:xi_final_app}
    \xi^\textrm{cr}=\frac{2}{9}.
\end{equation}
This is a remarkable result of Brazier's theory; it states that at the critical point of buckling, all tubes, independent of  thickness, radius or Young's modulus, become ovalized by the same ratio of $2/9$.

\section{The fundamental bandgap}\label{appendix:ZF}
The band structure of carbon nanotubes is based on graphene's band structure \cite{wallace} sliced-up  with lateral quantization lines. The exact wrapping of graphene into a nanotube
is determined by  the tube's chiral integers $(n,m)$ \cite{nanotube_book}. Now, any two integers can be related by other two integers $(q,p)$,  such that,
\begin{equation}\label{eq:n-m}
    n-m = 3q+p
\end{equation}
where $p$ takes one of the values ($0,+1,-1$). 

The  lateral $k$-vectors, $k_y$, must lie on quantization lines given by
\begin{equation}\label{eq:quantization_lines}
    k_y=\frac{\nu}{R},\,\,\,\,\,\,\,\textrm{where}\,\,\,\,\,\, \nu=0,\pm1,\pm2,\cdots.
\end{equation}

Graphene's Fermi surface  is found by the $k$-vectors that solve,
\begin{equation}\label{eq:wallace_hab_fundamental}
    \sum_{j=1}^3  \gamma_j e^{i\vec{k}\cdot\vec{l}_{j}}=0
\end{equation}
where  $\gamma_j$ are the three  overlap integrals of the nearest $\pi$-orbitals, and $l_j$ are their bond vectors, given by
\begin{eqnarray}\label{bonds}
\vec{l}_{1}&=&\frac{a^2}{4\pi R}\left((n+m)\hat{y}-\frac{1}{\sqrt{3}}(n-m)\hat{z}\right),\nonumber\\
\vec{l}_{2}&=&\frac{a^2}{4\pi R}\left(-m\hat{y}+\frac{1}{\sqrt{3}}(2n+m)\hat{z}\right),\\
\vec{l}_{3}&=&\frac{a^2}{4\pi R}\left(-n\hat{y}-\frac{1}{\sqrt{3}}(n+2m)\hat{z}\right),\nonumber
\end{eqnarray}
where $(n,m)$ are the chiral integers, and $(\hat{y},\hat{z})$ are the circumferential and axial coordinates, respectively.

Now under the assumption that graphene is isotropic, we can take all 
\begin{equation}\label{eq:equal_gamma}
    \gamma_j\equiv\gamma.
\end{equation} 
The solution of eq. (\ref{eq:wallace_hab_fundamental}) is then given by the two points,
\begin{eqnarray}\label{kfermi}
\vec{K}_{F1}&=&\frac{1}{3R}\left((m+2n)\hat{y}+m\sqrt{3}\hat{z}\right)\nonumber\\
\vec{K}_{F2}&=&\frac{1}{3R}\left((n-m)\hat{y}+(m+n)\sqrt{3}\hat{z}\right)\\ \nonumber
\end{eqnarray}
These are the Fermi points of graphene, in the nanotube's natural coordinates.

The spectrum near a Fermi point is linear, 
\begin{equation}\label{eg1}
     E=\pm\frac{\sqrt{3}}{2} a \gamma\,\delta k,
\end{equation}
where $E$ is the energy above the Fermi level and $\delta k$ is the distance from the nearest Fermi point.

Substituting (\ref{eq:n-m}) in (\ref{kfermi}) reveals that
 when $p=0$ they do lie on a quantization line (\ref{eq:quantization_lines}) -- these tubes are thus, according to zone-folding, metallic. For $p=\pm1$, on the other hand, the distance to the nearest quantization line is $1/3R$; these tube are thus semiconducting with an energy gap given by  substituting $1/3R$ in the linear spectrum (\ref{eg1}), 
\begin{equation}\label{eq:Eg_ZF}
    E_g=|p|\frac{a\gamma}{\sqrt{3} R}.
\end{equation}

\section{Fermi energy and  $C_s$}\label{appendix:fermi_and_C_s}
The lowering of the singlet ($S$) band below the conduction ($\pi$) band causes, as DFT simulations demonstrate  (table \ref{table:Ef}), a downshift not only of the bandgap but also the Fermi energy. While it is generally understood to be  a result of a large circumferential curvature of small tubes,  we wish to quantify it by extrapolating the published data for tubes in this radii range (regime $B$: section \ref{section:curvature_large}).

In this regime, the bandgap depends on the position of the singlet band above the Fermi points (eq. \ref{eq:Es_def}), which depends, in turn, on the proportionality factor $C_s$.  The DFT bandgaps of semiconducting, straight and un-deformed  tubes (table \ref{table:Eg}) in this regime ($R\lesssim4$\AA) give $C_s\approx 8 (eV\cdot$\,\AA$^2$). 




\begin{table}[htbp]
\begin{center}
\begin{tabular}{ l c c c} 
 \hline
 $\xi$ \,\,\,\,\,\,\,\,\,\,\,\,\,\,\,\,& $11,0$ & $10,0$ & $8,0$ \\ 
 \hline
 0 & 0.82 & 0.87 & 0.487\\
 0.0625 & 0.8 & 0.85 & 0.41\\
 0.125 & 0.75 & 0.825 & 0.3\\
 0.1825 & 0.65 & 0.7 & 0.075\\

 0.25 & 0.525 & 0.55 & 0\\
 0.3125 & 0.375 & 0.32 & 0\\
 0.375 & 0.225 & 0.075 & 0\\
 0.4375 & 0 & 0 & 0\\
 \hline
\end{tabular}
\end{center}\caption{Energy gaps (eV) in the respective tubes as a function of the deformation parameter $\xi$; data extracted from figure 1a in ref.  \cite{Shan_apl2005}}\label{table:deformation}
\end{table}

\begin{figure}[htbp]  
\begin{center}
\includegraphics[width=0.45\textwidth]{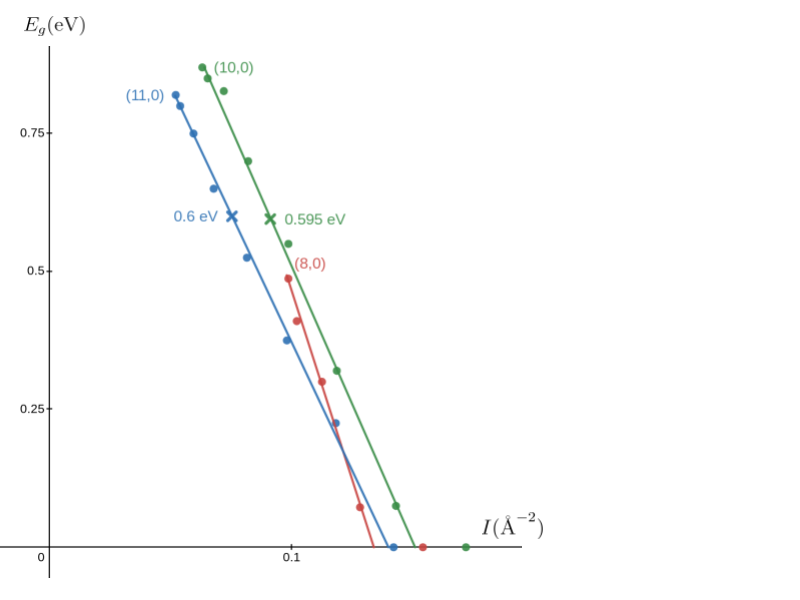}\caption{Bandgap vs. total circumferential curvature square  for the zigzags (11,0), (10,0) and (8,0).  Data points are given in table (\ref{table:deformation}); lines are plotted by eq. (\ref{eq:finall_gap}) (regime B) where $1/R^2$ was replaced by $I$ (see eq. \ref{eq:curv_square1}) with $C_s\sim8- 10$ (eV$\cdot$\AA$^2$). The  crosses label the critical points of buckling (given by eq. \ref{eq:cr_A2Rsquare}).
} \label{fig:gap_vs_curvature_square} 
\end{center}
\end{figure}

\begin{figure}[htbp]  
\begin{center}
\includegraphics[width=0.45\textwidth]{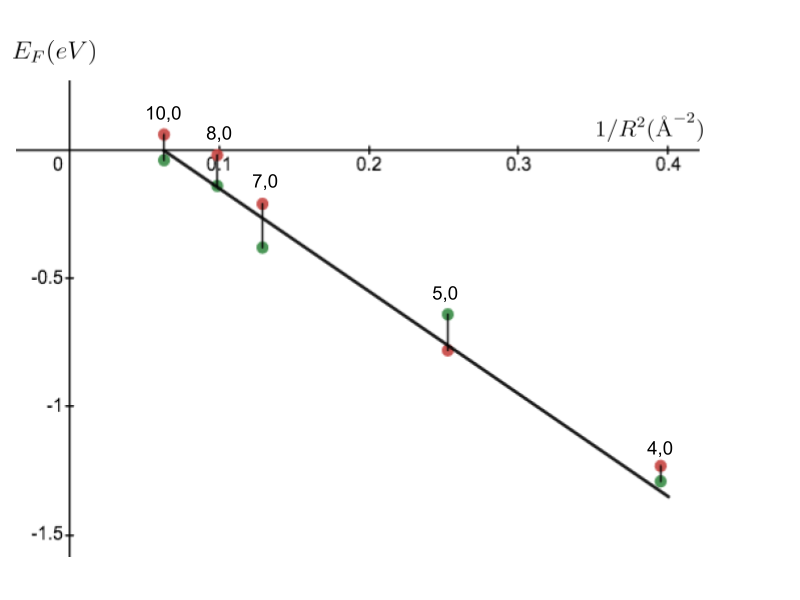}\caption{Fermi energy of a number of small zigzag  tubes vs. $1/R^2$. The line is a plot of eq. (\ref{eq:EF_simple}) with $C_s=8$\,(eV$\cdot$\AA$^2$).  Data is in table \ref{table:Ef} where red dots taken from ref. \cite{Barone2006}, green dots from \cite{Shan_prl}.
} \label{fig:fermi_vs_cur} 
\end{center}
\end{figure}

Moving to Fermi energy, DFT simulations
 show (table \ref{table:Ef}) that the Fermi-energy of tubes in this regime are  downshifted too in $\propto 1/R^2$ (fig. \ref{fig:fermi_vs_cur} ), while larger tubes have it  identical to graphene.
 The  $S$-band in this regime is the effective conduction band,  and thus,  the Fermi energy lies in the middle between it and the valence $\pi$-band (which is not shifted), 

\begin{equation}\label{eq:EF1}
    E_F=\frac{1}{2}(E_s+E_v)=\frac{1}{2}\left(E_s-\frac{E_g^\pi}{2}\right),
\end{equation}
 where the singlet band energy $E_s$ is given by eq. (\ref{eq:Es_def}) and the pure $\pi$ bandgap $E_g^\pi$ is given by eq. (\ref{eq:curvature_pi_gap}). This gives explicitly,

\begin{eqnarray}\label{eq:EF2}
     E_F&=&c_1-|p|\frac{\gamma a}{4\sqrt{3} R}\nonumber\\
     &-&\frac{\cos{3\alpha}}{R^2}\left(c_2+\textrm{sgn}[1-2p]\frac{\gamma a^2}{64}\right).
\end{eqnarray}
where the constants are
\begin{eqnarray*}
    c_1&=&\frac{E_g(10,0)}{4}+\frac{C_s}{2R_{10,0}^2}\approx0.53\, eV,\\
    c_2&=&\frac{C_s}{2}\approx4\, eV\cdot\textrm{\AA}^2,
\end{eqnarray*}
 where we used $E_g(10,0)=0.87\,eV$ (table \ref{table:Eg}), $R_{10,0}=4$\AA, and $C_s=8\,eV\cdot$\AA$^2$.


\begin{table}[htbp]
\begin{center}
\begin{tabular}{ c |  c  c } 
 \hline
 \,\,$n,m$\,\, &\,\, ref. \cite{Barone2006}\,\,& ref. \cite{Shan_prl}\,\,  \\ [0.5ex] 
 \hline
 $4,0$ &  $-1.23$ & $-1.29$ \\ [0.5ex]

 $5,0$ &  $-0.78$ & $-0.64$  \\ [0.5ex]

 $7,0$ &  $-0.21$ &  $-0.38$ \\ [0.5ex] 
 
 $8,0$ &  $-0.02$ &  $-0.14$ \\ [0.5ex] 
 
 $10,0$ &  $0.06$ & $-0.04$ \\ [0.5ex] 
 
 
 
 \hline
\end{tabular}
\end{center}
\caption{ Fermi  energies (in eV) of  semiconducting zigzag tubes. What was actually computed  are work-functions, $(WF)^\textrm{tube}$; Fermi energies shown here were then found relative to graphene  by  $E_F^\textrm{tube}=(WF)^\textrm{graphene}-(WF)^\textrm{tube}$, where    $(WF)^\textrm{graphene}=4.55$eV in ref. \cite{Barone2006} and $4.66$eV in ref.  \cite{Shan_prl}.} \label{table:Ef}
\end{table}

Now this can be further simplified if we neglect the second term in the parenthesis in eq. (\ref{eq:EF2}) and, since $E_F(R\geq R_c)=0$, following the downshift of the singlet band $E_s$ (eq. \ref{eq:Es_def}),
 \begin{equation}\label{eq:EF_simple}
 E_F=\frac{C_s}{2}\left(\frac{1}{R_c^2}-\frac{1}{R^2}\right)\cos{3\alpha},\,\,\,\,\,\,\,\,\,\,\,\,\,\, R\leq R_c,
 \end{equation}
where $R_c$ is given, by eq. (\ref{eq:Rc}); for zigzag tubes ($\alpha=0$), $R_c=R(10,0)=4$\AA. Eq. (\ref{eq:EF_simple}) is depicted in fig. (\ref{fig:fermi_vs_cur}) for semiconducting zigzag tubes in this range, together with DFT data. 
 
 Considering circumferentially ovalized tubes, following our procedure we first replace $1/R^2$ with $(1+9\xi^2/2)/R^2 \equiv I$ (eq. \ref{eq:curv_square1}),  where $\xi$ is the ovalization parameter. Explicitly,
 \begin{equation}\label{eq:EF_xi_final}
     E_F(\xi)=\frac{C_s}{2R^2}\left(\left(\frac{R}{R_c}\right)^2-1+\frac{9}{2}\xi^2\right)\cos{3\alpha},\,\,\,\,\, R\leq R_c.
 \end{equation}
 Comparing this with the DFT bandgaps (table \ref{table:deformation}), plotted in fig. (\ref{fig:gap_vs_curvature_square}), yields $C_s\sim8$ (eV$\cdot$\AA$^2$). 
 
 Finally, the fact that the analytic treatment in this work coincides with DFT on three different quantities: bandgaps of straight tubes, ovalized tubes, and Fermi energy of straight tubes, by having a single adjustable parameter, $C_s\sim8$ (eV$\cdot$\AA$^2$), is, in our opinion, a strong indication of the correctness of this treatment. 
\bibliography{kink_bib}
\end{document}